\documentclass[aps,prd,onecolumn,groupedaddress,showpacs,nofootinbib,amssymb]{revtex4}
\usepackage[dvips]{graphicx}
\usepackage{amssymb}
\usepackage{amsmath}
\usepackage{graphicx,,color}
\usepackage{amsfonts}
\usepackage{bm}
\usepackage{cancel}
\usepackage{comment}
\usepackage{floatflt}

\newcommand\be{\begin{equation}}
\newcommand\ee{\end{equation}}

\allowdisplaybreaks[4]

\begin{document}

\tolerance=5000

\title{$R^2$ Gravity Effects on the Kinetic Axion Phase Space}
\author{V.K. Oikonomou,$^{1,2}$}
\affiliation{$^{1)}$ Department of Physics, Aristotle University
of Thessaloniki, Thessaloniki 54124, Greece \\
$^{2)}$ L. N.
Gumilyov Eurasian National University, Nur-Sultan, Kazakhstan}
\email{voikonomou@auth.gr;v.k.oikonomou1979@gmail.com}


 \tolerance=5000

\begin{abstract}
In this work we consider the effect of an $R^2$ term on the
kinetic misalignment axion theory. By using the slow-roll
assumptions during inflation and the field equations, we construct
an autonomous dynamical system for the kinetic axion, including
the effects of the $R^2$ term and we solve numerically the
dynamical system. As we demonstrate, the pure kinetic axion
attractor is transposed to the right in the field phase space, and
it is no longer $(\phi,\dot{\phi})=(\langle \phi \rangle,0)$, but
it is $(\phi,\dot{\phi})=(\langle \phi '\rangle,0)$, with $\langle
\phi '\rangle\neq 0$ some non-zero value of the scalar field with
$\langle \phi '\rangle> \langle \phi \rangle$. This feature
indicates that the kinetic axion mechanism is enhanced, and the
axion oscillations are further delayed, compared with the pure
kinetic axion case. The phenomenological implications on the
duration of the inflationary era, on the commencing of the
reheating era and the reheating temperature, are also discussed.
\end{abstract}


\maketitle

\section{Introduction}

Particle dark matter is possibly the answer to the dark matter
problem, however to date, no dark matter particle has ever been
observed, see Refs.
\cite{Bertone:2004pz,Bergstrom:2000pn,Mambrini:2015sia,Profumo:2013yn,Hooper:2007qk,Oikonomou:2006mh}
for various particle dark matter theoretical searches. This is
possibly due to the fact that the dark matter particle has very
small mass. One theoretically appealing small mass candidate for
particle dark matter is the axion
\cite{Preskill:1982cy,Abbott:1982af,Dine:1982ah,Marsh:2015xka,Sikivie:2006ni,Raffelt:2006cw,Linde:1991km,Co:2019jts,Co:2020dya,Barman:2021rdr,Marsh:2017yvc,Odintsov:2019mlf,Nojiri:2019nar,Nojiri:2019riz,Odintsov:2019evb,Cicoli:2019ulk,Fukunaga:2019unq,Caputo:2019joi,maxim,Chang:2018rso,Irastorza:2018dyq,Anastassopoulos:2017ftl,Sikivie:2014lha,Sikivie:2010bq,Sikivie:2009qn,Caputo:2019tms,Masaki:2019ggg,Soda:2017sce,Soda:2017dsu,Aoki:2017ehb,Masaki:2017aea,Aoki:2016kwl,Obata:2016xcr,Aoki:2016mtn,Ikeda:2019fvj,Arvanitaki:2019rax,Arvanitaki:2016qwi,Arvanitaki:2014wva,Arvanitaki:2014dfa,Sen:2018cjt,Cardoso:2018tly,Rosa:2017ury,Yoshino:2013ofa,Machado:2019xuc,Korochkin:2019qpe,Chou:2019enw,Chang:2019tvx,Crisosto:2019fcj,Choi:2019jwx,Kavic:2019cgk,Blas:2019qqp,Guerra:2019srj,Tenkanen:2019xzn,Huang:2019rmc,Croon:2019iuh,Day:2019bbh,Odintsov:2020iui,Nojiri:2020pqr,Odintsov:2020nwm,Oikonomou:2020qah},
which is elusive and it is theorized that its mass may be smaller
than $m_a\leq 10^{-12}$eV, a fact that it is impressive. Unless
the LHC has large mass surprises for dark matter candidates, the
axion seems to be the last resort of particle dark matter. The
axion is a light scalar field, which naturally arises in string
theory as the string moduli.

In the recent literature, the terminology axion refers to an axion
like particle, but not to the QCD axion. The axion scalar has a
primordial pre-inflationary $U(1)$ Peccei-Quinn symmetry, which is
broken during inflation in the most popular axion models, which
are the canonical misalignment axion \cite{Marsh:2015xka} and the
kinetic misalignment axion
\cite{Co:2019jts,Co:2020dya,Barman:2021rdr}. In both the models
the $U(1)$ Peccei-Quinn symmetry is broken during the inflationary
era, and the axion obtains a large vacuum expectation and rolls to
its vacuum expectation value which is the minimum of the potential
The major difference between the two models is that in the case of
the canonical misalignment axion model, the axion has zero kinetic
energy initially, so when the axion reaches the minimum of the
potential, which is its vacuum expectation value, the axion
commences oscillations and thereafter redshifts as cold dark
matter. On the contrary, in the kinetic misalignment axion model,
the axion initially has a large kinetic energy, which actually
dominates over its potential. In effect, the axion rolls down to
its vacuum expectation value, but does not stop at the potential
minimum, which is also its non-zero vacuum expectation value, but
continues uphill deviating from its potential minimum. This
feature has a dramatic effect on the reheating era, since
basically the axion oscillations are significantly delayed
compared to canonical misalignment axion model.

In this paper we aim to investigate the effects of modified
gravity on the kinetic misalignment axion model. Modified gravity
\cite{reviews1,reviews2,reviews3,reviews4,reviews5} offers an
appealing theoretical framework in the context of which the
inflationary and the dark energy eras can be described in an
observationally viable and unified way
\cite{Nojiri:2003ft,Nojiri:2007as,Nojiri:2007cq,Cognola:2007zu,Nojiri:2006gh,Appleby:2007vb,Elizalde:2010ts,Odintsov:2020nwm,Sa:2020fvn},
and furthermore without having the shortcomings of the general
relativistic description of the dark energy era. Our aim is to
investigate the effects of a popular modified gravity model, that
of $R^2$ gravity, on the kinetic misalignment model. For our study
we shall adopt the dynamical systems approach, constructing an
autonomous dynamical system from the field equations of the
kinetic misalignment axion and we shall investigate in a
quantitative way, which is the final attractor of the dynamical
system.

By comparing the ordinary kinetic misalignment model with the
$R^2$-corrected kinetic misalignment axion model, we come to the
conclusion that the final attractor of the theory is different
from the pure kinetic misalignment axion case. Particularly, the
final attractor of the pure kinetic misalignment axion is
$(\phi,\dot{\phi})=(\langle \phi \rangle,0)$, with $\langle \phi
\rangle$ being the axion's vacuum expectation value during
inflation, however in the kinetic misalignment axion case, the
final attractor is $(\phi,\dot{\phi})=(\langle \phi '\rangle,0)$,
where $\langle \phi '\rangle\neq 0$ some non-zero value of the
scalar field with $\langle \phi '\rangle> \langle \phi \rangle$.
We show this feature numerically by studying the dynamical system,
and qualitatively this means that the axion does not settle to its
minimum of the potential, which is its vacuum expectation value
during inflation, but further continues its trajectory uphill
until it reaches the value $\langle \phi '\rangle\neq 0$. After
that it rolls down to the minimum of the potential, and the axion
commences its oscillations, when $\dot{\phi}\simeq V(\phi)$, and
it starts to redshift as cold dark matter.

Thus, the $R^2$ term further enhances the kinetic axion physics,
causing a larger delay for the start of the reheating era, a
feature that is phenomenologically important, since this delay is
basically an enhancement of the duration of the inflationary era.

\section{The $R^2$-corrected Kinetic Misalignment Axion Model and its Phase Space}

In this section we shall consider in a quantitative way the
effects of the $R^2$ term on the kinetic misalignment axion, by
using the phase space approach. Specifically we shall study the
dynamical system of the kinetic axion and by taking into account
the changes of the $R^2$ term on the dynamical system, we shall
quantitatively study the final attractor of the theory. A direct
comparison of the resulting phase space with the $R^2$-free model
shall also be taken into account. Before we proceed to our
analysis, let us briefly present the theoretical framework we
shall use, the field equations and let us describe the kinetic
misalignment axion mechanism.

We shall consider the following gravitational action,
\begin{equation}
\label{mainaction} \mathcal{S}=\int d^4x\sqrt{-g}\left[
\frac{1}{2\kappa^2}F(R)-\frac{1}{2}\partial^{\mu}\phi\partial_{\mu}\phi-V(\phi)
\right]\, ,
\end{equation}
with $\kappa^2=\frac{1}{8\pi G}=\frac{1}{M_p^2}$, and $G$ as usual
denotes Newton's gravitational constant. Also, $M_p$ denotes the
reduced Planck mass. For the purposes of this article, we shall
assume that the $F(R)$ gravity has the following form,

\begin{equation}\label{starobinsky}
F(R)=R+\frac{1}{M^2}R^2\, ,
\end{equation}
so it is basically the $R^2$ model. The parameter $M$ takes the
value $M= 1.5\times 10^{-5}\left(\frac{N}{50}\right)^{-1}M_p$, for
inflationary phenomenological reasons \cite{Appleby:2009uf}, with
$N$ denoting the $e$-foldings number as usual, but for the study
of the phase space of the cosmological system we shall use the
Planck units physical system. Considering a flat
Friedmann-Robertson-Walker (FRW) geometric background,
\begin{equation}
\label{metricfrw} ds^2 = - dt^2 + a(t)^2 \sum_{i=1,2,3}
\left(dx^i\right)^2\, ,
\end{equation}
the field equations corresponding to the action (\ref{mainaction})
are,
\begin{align}\label{eqnsofmkotion}
& 3 H^2F_R=\frac{RF_R-F}{2}-3H\dot{F}_R+\kappa^2\left(
\rho_r+\frac{1}{2}\dot{\phi}^2+V(\phi)\right)\, ,\\ \notag &
-2\dot{H}F=\kappa^2\dot{\phi}^2+\ddot{F}_R-H\dot{F}_R
+\frac{4\kappa^2}{3}\rho_r\, ,
\end{align}
\begin{equation}\label{scalareqnofmotion}
\ddot{\phi}+3H\dot{\phi}+V'(\phi)=0
\end{equation}
with $F_R=\frac{\partial F}{\partial R}$, while the ``dot''
denotes as usual differentiation with respect to the cosmic time,
while the ``prime'' differentiation with respect to the scalar
field, in our case the axion scalar field.

\begin{figure}
\centering
\includegraphics[width=18pc]{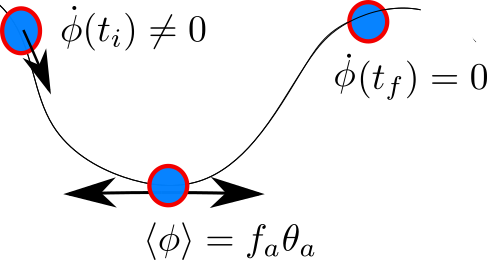}
\caption{A pictorial description of the kinetic misalignment axion
mechanism.}\label{plot2}
\end{figure}

Let us now describe in brief the kinetic axion mechanism in order
to better understand the new quantitative effects that the $R^2$
term brings along in the theory. For details on the kinetic axion
mechanism see for example
\cite{Co:2019jts,Co:2020dya,Barman:2021rdr}. In the context of the
kinetic axion mechanism, the axion primordially has an unbroken
$U(1)$ Peccei-Quinn symmetry, which is broken during the
inflationary era. Due to the broken symmetry, the axion acquires a
large vacuum expectation value $\langle \phi \rangle =\theta_a
f_a$, with $\theta_a$ being the initial misalignment angle, while
$f_a$ stands for the axion decay constant. During inflation, in
the context of the kinetic axion mechanism, the axion has a
non-zero and large kinetic energy. The kinetic axion mechanism is
pictorially described in Fig. \ref{plot2}.
Initially the axion has a small displacement from its vacuum
expectation value and a large kinetic energy. Due to the large
kinetic energy the axion does not stop to the minimum of the
potential, which is also its vacuum expectation value, but
continues uphill until it stops. Eventually the kinetic axion
rolls down again and when it reaches the minimum of the potential,
the axion oscillations occur. These oscillations make the axion
energy density to redshift as cold dark matter, and the axion
oscillations start when the reheating era commences. Primordially,
the axion has the following potential,

\begin{equation}\label{axionpotentnialfull}
V(\phi )=m_a^2f_a^2\left(1-\cos (\frac{\phi}{f_a}) \right)\, ,
\end{equation}
however, during inflation, the axion potential for small
displacements around its vacuum expectation value is,
\begin{equation}\label{axionpotential}
V(\phi )\simeq \frac{1}{2}m_a^2\phi^2\, ,
\end{equation}
an approximation which holds true for $\phi\ll f_a$ or
equivalently for $\phi\ll \langle \phi \rangle $. Thus
essentially, from a dynamical point of view, the final attractor
of the kinetic axion is its vacuum expectation value, at which
point the axion oscillations commence.

Now let us form an autonomous dynamical system for the
$R^2$-corrected kinetic axion in order to see quantitatively the
effects of the $R^2$ term on the phase space of the axion. For the
study of the dynamical system, we shall adopt the Planck units
physical system in which $\hbar=c=\kappa=1$, recall $\kappa=1/M_p$
so basically $M_p=1$.

\begin{figure}
\centering
\includegraphics[width=18pc]{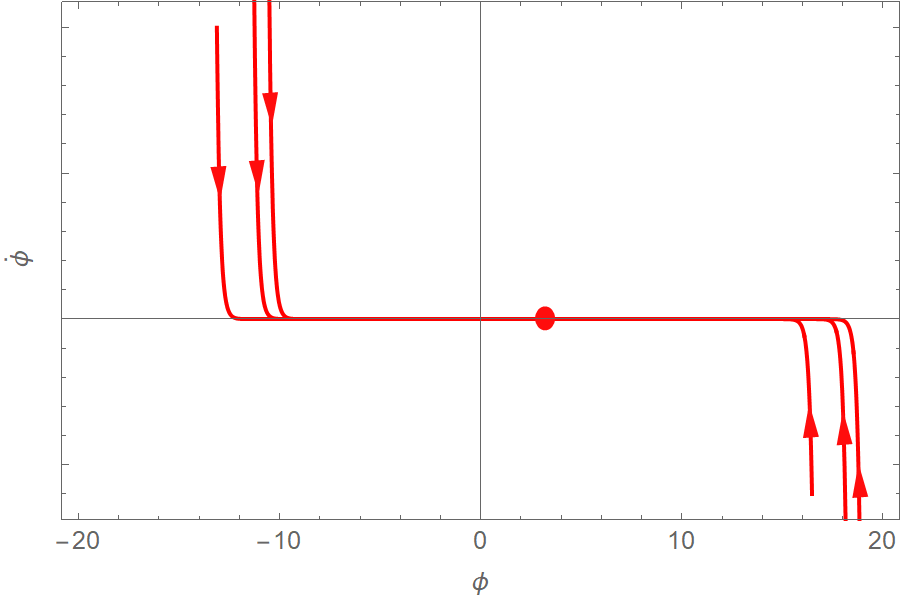}
\includegraphics[width=18pc]{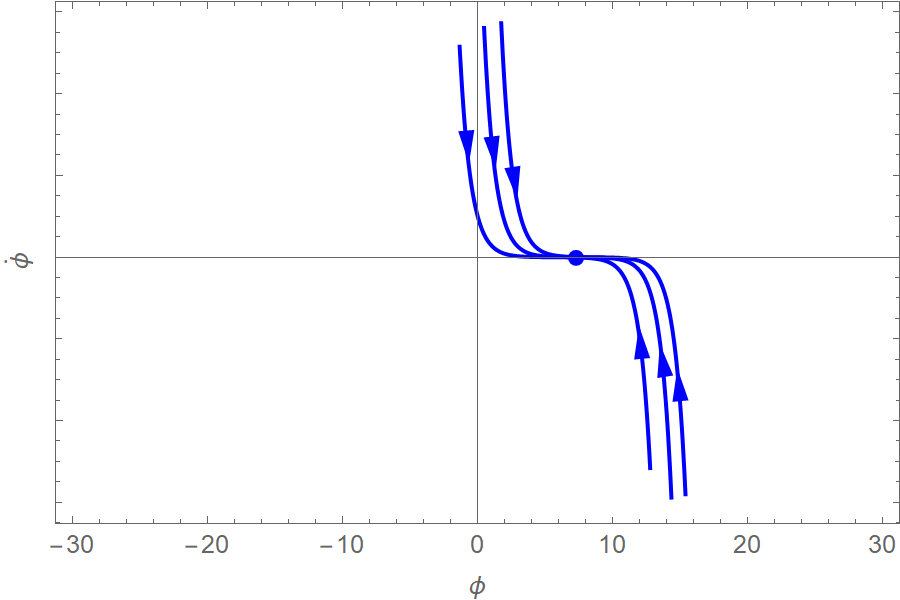}
\caption{The phase space attractors in the case of the pure
kinetic axion theory (left panel) and the $R^2$-corrected kinetic
axion case (right panel) for various initial conditions quantified
in terms of a non-zero initial value of
$\dot{\phi}$.}\label{kinetic}
\end{figure}

Let us consider the field equations (\ref{eqnsofmkotion}) and
(\ref{scalareqnofmotion}), and in the slow-roll approximation for
the $R^2$ model, the Friedmann equation takes the form,
\begin{equation}\label{freidmaneqn}
3 H^2\simeq
-3H^2\frac{\dot{H}}{M^2}+\kappa^2V+\kappa^2\dot{\phi}^2\, .
\end{equation}
while the Raychaudhuri equation takes the following form at
leading order,
\begin{equation}\label{raychaunhduhri}
-2\dot{H}-2\frac{\dot{H}^2}{M^2}\simeq \kappa^2\dot{\phi}^2\, .
\end{equation}
Upon solving the Raychaudhuri equation algebraically in terms of
$\dot{H}$, we obtain,
\begin{equation}\label{doth}
\dot{H}=\frac{1}{2} \left(M \sqrt{M^2-2 \dot{\phi}^2 \kappa
^2}-M^2\right) \, ,
\end{equation}
hence, upon substituting $\dot{H}$ from Eq. (\ref{doth}) in the
Friedmann equation (\ref{freidmaneqn}), the Hubble rate reads,
\begin{equation}\label{hubbleratefinal}
H=\frac{\kappa  \sqrt{\dot{\phi}^2+V}}{\sqrt{\frac{3 \sqrt{M^2-2
\dot{\phi}^2 \kappa ^2}}{2 M}+\frac{3}{2}}} \, .
\end{equation}
Upon substituting the Hubble rate from Eq. (\ref{hubbleratefinal})
into Eq. (\ref{scalareqnofmotion}), and by introducing the
variable $\varphi=\dot{\phi}$ and using the potential of Eq.
(\ref{axionpotential}) which is valid during inflation, we obtain
the following dynamical system,
\begin{equation}\label{dynamicalsystemfinal}
\frac{d \varphi}{d\phi}\varphi=  -3 \kappa \varphi \frac{
\sqrt{\varphi^2+V}}{\sqrt{\frac{3 \sqrt{M^2-2 \varphi^2 \kappa
^2}}{2 M}+\frac{3}{2}}}-m^2\phi \, ,
\end{equation}
We can solve numerically the dynamical system
(\ref{dynamicalsystemfinal}) by using various initial conditions
for $\dot{\phi}$ at $t=0$, making sure though that
$\dot{\phi}(t=0)\neq 0$. The results of our numerical analysis can
be found in Fig. \ref{kinetic}. In the left plot of Fig.
\ref{kinetic} we present the kinetic axion phase space attractor
in the absence of the $R^2$ term, while in the right plot we
present the $R^2$-corrected kinetic axion phase space attractor.
In both plots, the two axes of symmetry meet at the kinetic axion
phase space attractor in the absence of the $R^2$ term. As it can
be seen in the left plot, the new attractor of the theory is not
$(\phi,\dot{\phi})=(\langle \phi \rangle,0)$, but it is
$(\phi,\dot{\phi})=(\langle \phi '\rangle,0)$, with $\langle \phi
'\rangle\neq 0$ some non-zero value of the scalar field with
$\langle \phi '\rangle> \langle \phi \rangle$.
Thus in the $R^2$-corrected kinetic axion case, the kinetic axion
mechanism is enhanced, and the reheating era starts later than in
the pure kinetic axion theory. This is due to the fact that the
cosmological system is attracted to the attractor
$(\phi,\dot{\phi})=(\langle \phi '\rangle,0)$, thus the kinetic
axion further delays its downhill motion to the vacuum expectation
value, and it basically starts the downhill motion to its vacuum
expectation value much more later than the pure kinetic axion.
This effect indicates that the reheating era in the
$R^2$-corrected kinetic axion case starts at a much more later
time compared to the pure kinetic axion case. In effect, in the
combined $R^2$-corrected kinetic axion theory, the inflationary
era is somewhat prolonged. This feature is also pointed out in
Ref. \cite{submitted}, and has some quantitative phenomenological
implications. We need to note that the various phase space plots
of the two panels in Fig. \ref{kinetic} correspond to different
initial conditions on the parameter $\varphi (t)$ at $t=0$, and
recall that $\varphi (t)=\dot{\phi}$. Thus giving various initial
velocities we get different curves in each panel, but for the
$R^2$ kinetic axion curves, which are the blue ones, the attractor
point is shifted to the right. Since solving analytically the
dynamical system \ref{dynamicalsystemfinal} is a formidable task,
we limit ourselves to a qualitative approach, which shows the
shift caused by the $R^2$ gravity term. We do not discuss the
actual numerical values of the shifted plot, since we are working
in Planck units.

\section{Conclusions}

In this work we investigated quantitatively by using a phase space
approach, the effects of an $R^2$ term on the dynamical evolution
of the kinetic axion. Specifically, by using solely the slow-roll
assumptions, we constructed an autonomous dynamical system for the
kinetic axion, including the $R^2$ effects. By using several
appropriate initial conditions for the initial velocity of the
scalar field $\dot{\phi}$ we solved numerically the dynamical
system and we studied the phase space attractors for both the
$R^2$-corrected kinetic axion model and for the pure kinetic axion
model. As we demonstrated, the pure kinetic axion attractor is
transposed to the right in the phase space plot, and it is no
longer $(\phi,\dot{\phi})=(\langle \phi \rangle,0)$, but it is
$(\phi,\dot{\phi})=(\langle \phi '\rangle,0)$, with $\langle \phi
'\rangle\neq 0$ some non-zero value of the scalar field with
$\langle \phi '\rangle> \langle \phi \rangle$. This effect causes
a further delay on the start of the axion oscillations in the case
of the $R^2$-corrected axion case, thus the reheating era is
somewhat postponed. The same conclusion is found in Ref.
\cite{submitted}, and in fact, as it is pointed out in Ref.
\cite{submitted}, the $R^2$ inflationary era is prolonged due to
the presence of the kinetic axion. Thus the combined effect of the
$R^2$ term and the kinetic axion is to prolong the inflationary
era and the reheating era starts somewhat later compared to the
pure $R^2$ model and the pure kinetic axion model. This can have
phenomenological consequences even on the inflationary era and the
reheating temperature. The former issue is addressed in Ref.
\cite{submitted}, the latter is deferred to a future work.

\section*{Acknowledgments}

This research is funded by the Committee of Science of the
Ministry of Education and Science of the Republic of Kazakhstan
(Grant No. AP14869238) (V.K.O).

\end{document}